\documentclass[twocolumn,showpacs,preprintnumbers,amsmath,amssymb]{revtex4}

\usepackage{graphicx}
\usepackage{dcolumn}
\usepackage{bm}
\usepackage{color}

\begin{document}


\title{Fast sensing of double-dot charge arrangement and\\  spin state with an rf sensor quantum dot}

\date{\today}

\author{C. Barthel$^1$}
\author{M. Kj\ae{}rgaard$^{1,2}$}
\author{J. Medford$^{1}$}
\author{M. Stopa$^1$}
\author{C. M. Marcus$^1$}
\author{M. P. Hanson$^3$}
\author{A. C. Gossard$^3$}
\affiliation{$^1$Department of Physics, Harvard University,
Cambridge, Massachusetts 02138, USA\\
$^2$Nano-Science Center, Niels Bohr Institute, University of Copenhagen, Universitetsparken 5, DK-2100 Copenhagen, Denmark\\
$^3$Materials Department, University of California, Santa Barbara, California 93106, USA}

\begin{abstract}
Single-shot measurement of the charge arrangement and spin state of a double quantum dot are reported, with 
 times down to 100 ns. Sensing uses radio-frequency reflectometry of a proximal quantum dot in the Coulomb blockade regime. The sensor quantum dot is up to 30 times more sensitive than a comparable quantum point contact sensor, and yields three times greater signal to noise in rf single-shot measurements. Numerical modeling is qualitatively consistent with experiment and shows that the improved sensitivity of the sensor quantum dot results from reduced screening and smaller characteristic energy needed to change transmission.

\end{abstract}


\maketitle

Experiments on few-electron quantum dots \cite{Hanson07}, including spin qubits, have benefitted in recent years from the use of proximal charge sensing, a technique that allows the number and arrangement of charges confined in nanostructures to be measured via changes in conductance of a nearby sensor to which the device of interest is capacitively coupled  \cite{Field93,Ihn09}. Quantum point
contacts (QPCs) have been widely used as charge sensors, allowing, for instance, high-fidelity single-shot readout of spin qubits via spin-to-charge conversion \cite{elzerman04b,Singleshotpaper}. Single electron
transistors (SETs) based on metallic tunnel junctions, and gate defined sensor quantum dots~(SQD), conceptually equivalent to SETs, have also been widely used as proximal sensors, and provide similar sensitivity and bandwidth \cite{schoelkopf98, Buehler05,Lu03,Fujisawa05}.  As a typical application, measuring the state of a spin qubit via spin-to-charge conversion involves determining whether two electrons in a double quantum dot are in the $(1,1)$ or the $(0,2)$ charge configuration, where (left, right) denotes occupancies in the double dot [Fig.~1(a)], on time scales faster than the spin relaxation time \cite{Singleshotpaper}. 

In this Communication, we demonstrate the use of a sensor quantum dot for fast charge and two-electron spin-state measurement in a GaAs double quantum dot, biased near the (1,1)-(0,2) charge transition. We compare the performance of the SQD to conventional quantum point contact (QPC) sensors for dc and radio-frequency (rf) measurement. We find experimentally that the SQD is up to 30 times more sensitive, and provides roughly three times the signal to noise ratio (SNR) of a comparable QPC sensor for detecting the charge arrangement and spin state of a double quantum dot. Numerical simulations, also presented, give results consistent with experiment and elucidate the role of screening in determining the sensitivity of these proximal charge sensors. 

\begin{figure}[b]
\includegraphics[width=3.2 in] {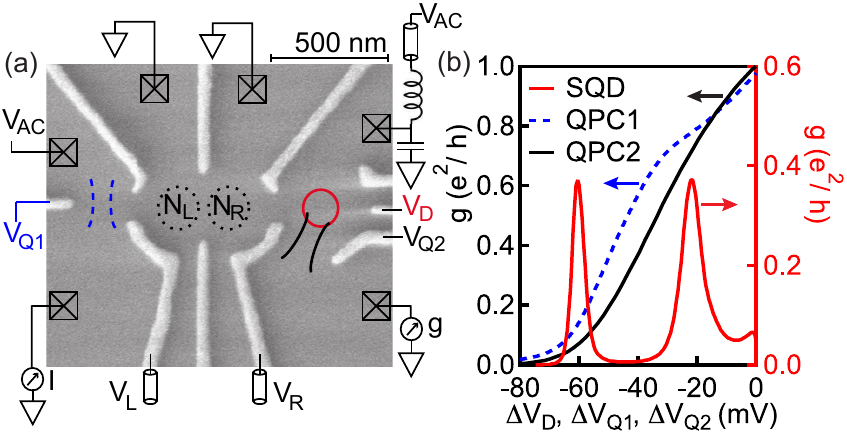}
\caption{\label{Fig1} (Color online) (a) Micrograph of lithographically identical device. Gate voltages
$V_{\rm{L}}$ and $V_{\rm{R}}$
control the double dot charge state $(N_{\rm{L}},N_{\rm{R}})$ (see Fig.~\ref{Fig2}).  Quantum point contact QPC1 (blue, dashed) controlled by gate voltage $V_{{\rm{Q1}}}$, sensor quantum dot~(SQD) (red, solid), by plunger gate $V_{{\rm{D}}}$, can also be operated as a point contact (QPC2) (black, solid) apply gate voltage $V_{{\rm{Q2}}}$ to the bottom gate with top two grounded. QPC1(2) and SQD measured by dc transport in first device.  SQD measured by rf reflectometry in subsequent cooldown of second identical device. (b) DC conductance, $g$, of
QPC1,2 (left scale) and SQD (right scale) as a function of gate voltage changes
$\Delta V_{{\rm{D}}}$, $\Delta V_{{\rm{Q1}}}$, and $\Delta V_{{\rm{Q2}}}$.
}
\end{figure}

Double quantum dots with integrated sensors are defined by Ti/Au depletion gates on
a GaAs/Al$_{0.3}$Ga$_{0.7}$As heterostructure with a two-dimensional
electron gas  (density $2\times10^{15} ~\rm{m}^{-2}$, mobility
$20~\rm{m}^2$/Vs) 100 nm below the surface. The charge state of the double quantum dot is controlled by gate voltages $V_{\rm{L}}$, $V_{\rm{R}}$ [see Fig. 1(a)]. Three gates next to the right dot  form the SQD, which is operated in the multi-electron Coulomb blockade (CB) regime, with center gate voltage $V_{\rm{D}}$ setting the SQD energy. A single gate next to the left dot forms a QPC sensor (denoted QPC1) whose conductance is controlled by gate voltage $V_{\rm{Q1}}$. A second QPC sensor (QPC2) results when the center and top gate voltages of the SQD are set to zero, with only the bottom gate set to $V_{\rm{Q2}}$.

Measurements were carried out in a dilution refrigerator at electron temperature $\sim 150$~mK, configured for high-bandwidth gating, rf reflectometry and low-frequency (dc) transport.  
Low-frequency conductance was measured using a voltage bias of $\sim50~\mu$V  at 197 Hz with a lock-in time constant of 100 ms. Two nominally identical devices were measured and showed similar behavior. In the first device, dc sensing was measured in QPC1, QPC2 and the SQD, along with single-shot rf reflectometry data for QPC1. The single-shot data for QPC1 in this device was discussed in detail in Ref.~\cite{Singleshotpaper}. In the second device, single-shot rf reflectometry \cite{Singleshotpaper} for the SQD was measured.

\begin{figure}
\includegraphics[width=3.2 in] {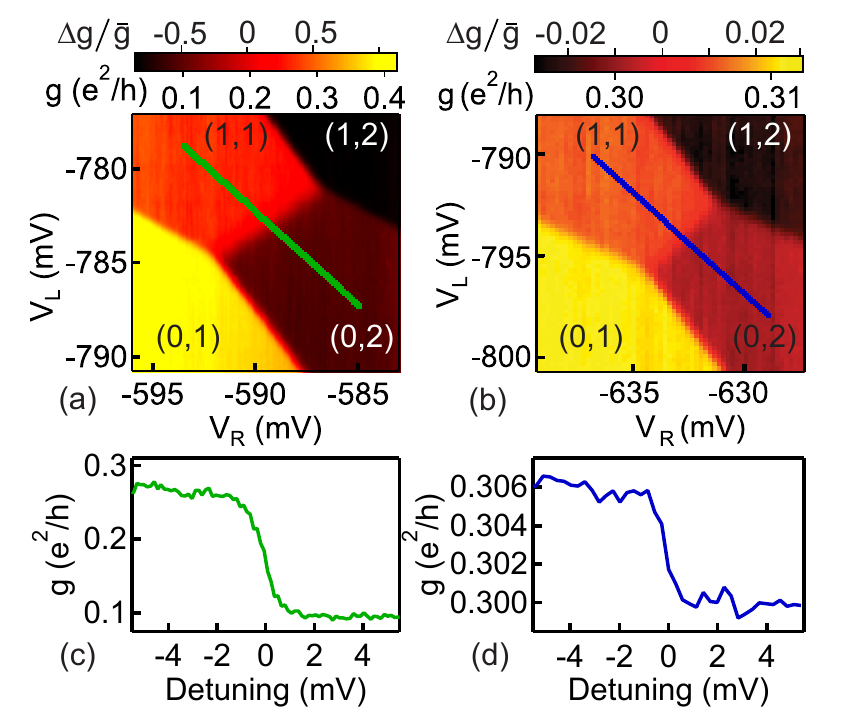}
\caption{\label{Fig2} (Color online) (a) Sensor quantum dot (SQD) dc conductance $g$ as a function of voltages
$V_{\rm{L}}$ and $V_{\rm{R}}$,
with charge occupancy $(N_{\rm{L}},N_{\rm{R}})$ indicated. Note the large relative change in conductance, $\Delta g / \bar{g} \sim 0.9$ as double dot switches from (0,2) to (1,1). (b) QPC2 conductance shows a small ($\Delta g / \bar{g}\sim 3\%$) change as the dot switches from (0,2) to (1,1). A similar value is seen for QPC1 (not shown).  (c, d) Cuts through (a), (b) respectively. All data is for Device 1.}
\end{figure}

Conductance of the SQD shows CB peaks as a function of plunger gate voltage, $\Delta V_{\rm{D}}$, while  conductances of QPC1 and QPC2 decrease smoothly, with  $\sim 10 \times$ lower maximum slope, as gate voltages $\Delta V_{\rm{Q1(Q2)}}$ are made more negative [Fig.~1(b)].  The greater slope of the SQD conductance versus gate voltage, compared to the QPC, is closely related to its higher sensitivity as a charge sensor (though not quantitatively, as lever arms to gates and dots differ). With $V_{\rm{D}}$ set on the side-wall of a CB peak, dc conductance of the SQD as a function of $V_{\rm{L}}$
and $V_{\rm{R}}$ indicates the charge state of the double dot [Fig.~2(a)]. Parasitic gating
of the SQD is compensated by trimming $V_{\rm{D}}$ and
$V_{\rm{Q2}}$ as  $V_{\rm{L}}$ and $V_{\rm{R}}$ are swept, to keep the
SQD conductance roughly constant on plateaus of fixed double dot charge arrangement. SQD conductance, centered around $\bar{g}=(g_{(1,1)}+g_{(0,2)})/2\sim 0.2~\rm{e}^2/\rm{h}$, changes by $\Delta g_{\rm{SQD}} \sim 0.2~\rm{e}^2/\rm{h}$ when the double dot charge arrangement changes from $(0,2)$ to $(1,1)$ (Fig.~2(c),.  Figure 2(d) shows corresponding~\cite{shiftdisclaimer} changes in QPC2 conductance, which changes by $\Delta g_{\rm{QPC2}}\sim 0.01~\rm{e}^2/\rm{h}$ around $\bar{g}\sim0.3~\rm{e}^2/\rm{h}$ for the same charge rearrangement, consistent with values in the literature. The ratio of conductance changes, $\Delta g_{\rm SQD}/ \Delta g_{\rm QPC2} \sim 30$, is a measure of the relative sensitivity of SQD and QPC2 to the double dot charge state. 

To demonstrate fast measurement of a spin qubit via spin-to-charge conversion, the SQD is configured as the resistive element in an rf reflectometry circuit~\cite{Reillyapl07}, following Ref.~\cite{Singleshotpaper}, and biased via $V_D$ on the sidewall of a CB peak. The reflected rf amplitude, $v_{\rm{rf}}$, tracks SQD conductance. Gate pulses applied to $V_{\rm{L}}$ and $V_{\rm{R}}$ first prepare the ground state singlet in $(0,2)$, then separate the spins by moving to point S, deep in $(1,1)$, for a time $\tau_{\rm{S}} = 1 - 200$~ns~\cite{tsstepping}, allowing precession between $(1,1)S$ and $(1,1)T_0$ driven by hyperfine fields, then move to the measurement point M in (0,2) for $\tau_{\rm{M}}^{\rm{max}}=5~\mu$s  (Fig.~2(a) and Ref.~\cite{Singleshotpaper}). At M, only the singlet configuration of the two spins can rapidly move to the (0,2) ground state; spin triplets remain trapped in $(1,1)$ for the spin relaxation time~\cite{Singleshotpaper}.

With rf excitation applied to the SQD only during the measurement interval at point M, the reflectometry signal, $v_{\rm{rf}}$, is digitally integrated over a subinterval of duration $\tau_{\rm{M}}$ to yield a
single-shot measurement outcome $V_{\rm{rf}}$. From histograms of
$3\times10^4$ $V_{\rm{rf}}$ measurements (with $0.7$~mV binning), probabilities, $P$, of single-shot outcomes can be estimated for each value of $\tau_{\rm{M}}$. As seen in Figs.~3(a,b), measurement noise decreases with increasing integration time, allowing distinct peaks---indicating singlet [i.e., (0,2)] and triplet [i.e., (1,1)] outcomes---to be distinguished for $\tau_{\rm M} > 100$ ns. The difference between singlet and triplet output voltages, the signal $\Delta V = V_{\rm{rf}}^{T} - V_{\rm{rf}}^{S}$ reflects the rf sensitivity of the SQD to the single-charge motion from (1,1) to (0,2). 

Experimental $P(V_{\rm{rf}})$ curves for the SQD are in good agreement with theoretical models~\cite{Singleshotpaper}, as shown in Fig.~3(b).  Fits of the model give values for  the spin relaxation time, $T_1 =13~\mu$s, the mean triplet probability, $\langle
P_{T}\rangle = 0.46$, the peak width,  $\sigma_{\rm{rf}}$, and the peak positions, $V_{\rm{rf}}^{S}$  and $V_{\rm{rf}}^{T}$. The resulting signal to noise ratio, SNR $=\Delta V/\sigma_{\rm rf} $ is shown in Fig.~3(c) as a function of $\tau_{\rm{M}}$, along with the SNR for QPC1. A direct comparison must take into account that the SQD data in all panels of Fig.~3 used -99 dBm applied rf power ($\sim 0.15~$mV), while the QPC1 data in Fig.~3(c) used -89 dBm applied rf power ($\sim 0.45~$mV), values chosen to maximize the SNR for each. For both QPC1 and SQD, the output signal $\Delta V$ saturated at higher powers, due in part to broadening of the conductance features due to heating and finite bias.

SNR for both the SQD and QPC1 improve with increasing integration time, as shown in Fig.~3(c). Fitting the measured SQD signal to noise ratios to the form $=\Delta V/\sigma_{\rm rf} $ , with $\sigma_{\rm{rf}} = \sigma_0 \sqrt{1\mu{\rm{s}} / (\tau_{\rm{M}}+\tau_0)}$, yields an intrinsic integration time, $\tau_0=190$~ns, due to the $\sim1.5$~MHz bandwidth of the reflectometry circuit, a signal, $\Delta V_{\rm{SQD}} = 33$~mV, and a characteristic width, $\sigma_0=5$~mV, the measurement noise for one microsecond total integration time. The ratio $\Delta V/ \sigma_0 $ represents a characteristic SNR, which is 6.6 for this SQD. A similar measurement of the characteristic SNR for QPC1, at 10 dB higher applied rf power, yields a value  2.2~\cite{Singleshotparams}, with $\Delta V_{\rm{QPC1}} = 10$~mV and $\sigma_0=4.5$~mV.  

For both SQD and QPC1, analysis \cite{KorotkovAPL99} predicts signals, $\Delta V$, consistent with measured values, and widths, $\sigma_0$, due to shot noise that are considerably lower than the measured peak widths. Specifically, $\sigma_0\sim 1.5 (3)$~mV and $\sim 3$ mV are expected for SQD (QPC1) \cite{Johnsonnoise}. This is roughly one tenth (half) of the total noise for the SQD (QPC1). The remaining measurement noise for both sensors is due to charge-, gate- and instrumentation noise, predominantly from the cryogenic amplifier~\cite{Reillyapl07,Johnsonnoise}. We conclude, based on the single-shot data, that the measured SQD offers improved SNR compared to a a comparable QPC sensor, ${\rm SNR}_{\rm SQD}/{\rm SNR}_{\rm QPC1} \sim 3$. The improvement is not as large as the relative improvement in sensitivity at dc, $  \Delta g_{\rm{SQD}} / \Delta g_{\rm{QPC1}} \sim 10$, mainly due to a lower rf power saturation of the SQD SNR and the experimental noise floor of the measurement setup.

\begin{figure}
\includegraphics[width=3.2 in]{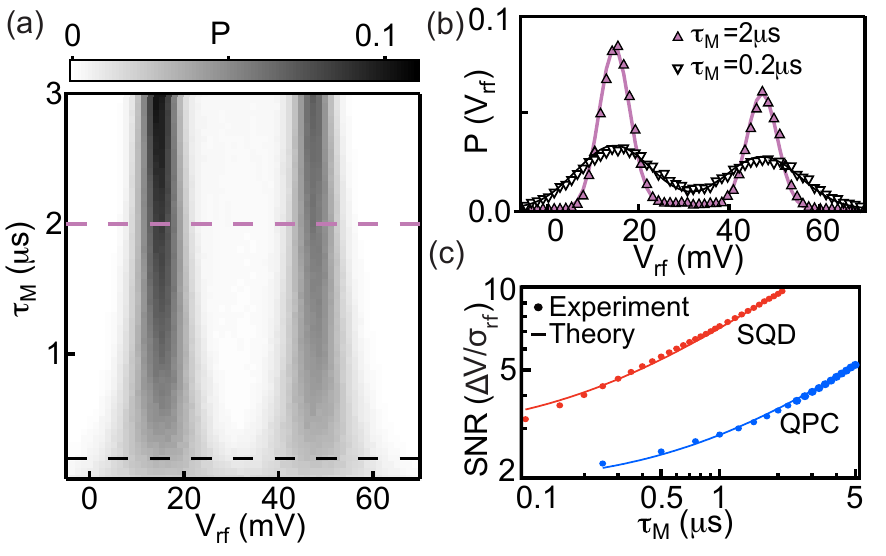}
\caption{\label{Fig3} (Color online) (a) Probability density $P$ of single-shot outcomes,
$V_{\rm{rf}}$, ($0.7$~mV binning) as a function of integration time, $\tau_M$, of the rf - charge signal $v_{\rm{rf}}$ \cite{tsstepping}. Measured with SQD, Device 2. The left
[right] peak corresponds to the (0,2) [(1,1)] charge state and
therefore singlet [triplet] measurement
outcomes~\cite{Singleshotpaper}.  (b) Cuts of $V_{\rm{rf}}$ along the dashed lines in (a),
along with theoretical curves \cite{Singleshotpaper}. (c) Signal to noise ratio (SNR), defined as peak separation $\Delta V$ divided by peak width $\sigma$ as a function of measurement integration time $\tau_M$ for QPC1 (Device 1, -89 dBm rf power) and SQD (Device 2, -99 dBm rf power), along with theory curves (see text).}
\end{figure}

To investigate QPC and SQD performance numerically, we consider the sensitivity, $s$, as the change in conductance in response to a change in voltage, either applied to a gate or arising from a charge rearrangement. Modeling the specific device geometry, 
for QPC1, $s_{\text{QPC}} \equiv  \frac{\partial
g}{\partial V_{\text{Q1}}} =  \frac{\partial g}{\partial
\phi_{\text{SP}}} \frac{\partial \phi_{\text{SP}}}{\partial
V_{\text{Q1}}}$, where $\phi_{\text{SP}}$ is the electrostatic
potential at the saddle point of QPC1. For the SQD, $s_{\text{SQD}}
\equiv \frac{\partial g}{\partial V_{\text{D}}} =
\frac{\partial g}{\partial
\phi_{\text{\text{dot}}}}\frac{\partial
\phi_{\text{\text{dot}}}}{\partial V_{\text{D}}}$, where
$\phi_{\text{\text{dot}}}$ is the electrostatic potential in the center of the SQD.

 For the QPC, the conductance, $g$, and its derivative with respect to potential, is
calculated as a thermal average over the transmission probability, following Ref.\,
\cite{Fertig87}. The width of the riser between conductance
plateaus scales as ${\cal E} \equiv \sqrt{\hbar^2 U_x/2 m}$, where
$U_x$ is the curvature of the saddle potential in the direction of
the current. The self-consistent calculation presented below yields
${\cal E} \sim 0.2$ meV, an order of magnitude greater
than $k_B T$. Thus the riser width is roughly independent of temperature.
The SQD conductance is modeled by a master equation \cite{Beenakker} assuming transmission via a single orbital level in the dot. This approach is applicable, given the single-particle level spacing is large, $\sim 200~\mu$eV, but is only valid for small tunneling rate, $\gamma$, from the dot to the leads, such that $\hbar \gamma \ll k_B T$. In the experiment, a larger coupling is used, such that $\hbar \gamma \sim k_B T$. This gives rise to some quantitative discrepancy between the model and the experiment, but the qualitative comparison between SQD and QPC performance remains valid. 

The lever-arm terms in the definitions of sensitivity, $\partial \phi /\partial V_D$ for the SQD and $\partial \phi /\partial V_{Q1(2)}$ for the QPCs, depend on positions of nearby conductors that screen the interaction between source of the voltage and the potential at the
target point. For QPC1(2), a change of $V_{\text{Q1(Q2)}}$ is
screened as charge in the leads of the QPC flow in or out of the saddle
region and \emph{opposes} the change
of $\phi_{\text{SP}}$ caused by the gate voltage change. In contrast, the SQD lever arm is primarily
determined by screening from other gates, rather than the 2DEG itself because the dot is
isolated by tunnel barriers and the charge is fixed by CB. Numerical calculation below gives a lever arm that is typically $\sim 20$ times greater for an SQD than for a QPC. Thus 2DEG screening
substantially influences  sensor response.

Conductances of the SQD and QPCs are calculated using the SETE code~\cite{NNINC,Stopa96}, which simulates the 3D electronic structure of the device within the effective-mass local-density-approximation to density functional theory. The calculation
produces the total free energy of the SQD as a function of $V_D$ and
$N$, enabling a calculation of the conductance in the single-level CB
regime~\cite{Stopa93}. Figure 4(a) shows a plot of the
calculated SQD conductances, and their difference, between the cases
where the double dot charge is held in the (0,2) and (1,1) states,
as a function of gate voltage offset $\Delta V_D$. For this calculation, the ratio $\hbar \gamma / k_B T$ is set to unity, based on experimental peak conductance values  [Fig.~1(b)]. We note, however, that the fractional change of conductance, $\Delta g/\bar{g}$, across the transition from (0,2) to (1,1) does not depend on $\hbar \gamma / k_B T$. For
QPC1, the evolution of the potential profile with varying
$V_{Q1}$ is calculated with SETE. The (1,1) and (0,2) conductances in Fig.~4(b)
are evaluated by solving the transverse Schr\"{o}dinger equation in
slices through the QPC and evaluating a 1D WKB expression for the
transmission.

\begin{figure}[h!]
\includegraphics[width=3.2in]{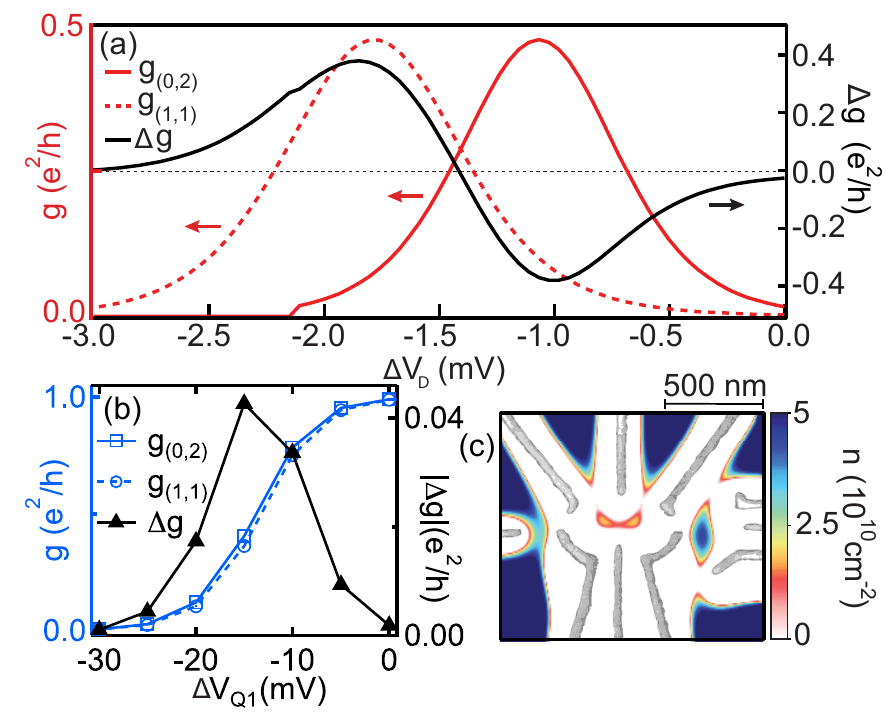}
\caption{ (Color online) (a) Conductance (simulation) through the SQD as function of
the gate voltage $V_{D}$, for the two double dot charge-states $(1,1)$ and
$(0,2)$. (b) Conductance through QPC1 at 5
mV intervals (lines are guide to the eye). (c) Electron density
profile for typical gate voltages in the (1,1) configuration, with superimposed micrograph of device. The color scale is 
centered near $2.5 \times 10^{10} \text{cm}^{-2}$ to accentuate the
charge in the dots and the saddle point of QPC1.} \label{fig:numcond}
\end{figure}

In the experiment, $V_D$ and $V_{\text{Q1}}$ are not swept,
rather they are held at their most sensitive point and the
conductance (through QPC or SQD) is allowed to change due to the
change in double dot state. The most sensitive points of the sensors are at the extrema of $\Delta g$. Here, the ratio $|\Delta g| /
\bar{g}$ is $\sim 1.4$ for the SQD and $\sim 0.1$ for QPC1, roughly consistent with experiment.

A color scale plot of the 2D electron density for typical gate
voltages is shown in Fig.~4(c). 

In conclusion, by taking advantage of the increased sensitivity and SNR of
a sensor quantum dot in the CB regime (compared to a proximal QPC), we have demonstrated single-shot spin-to-charge readout of a few-electron double quantum dot in $\sim 100\,$ns with $SNR \sim 3$ (Fig.~3), representing an order of magnitude improvement over previous results~\cite{Singleshotpaper}. Numerical simulation based on density functional theory yields good qualitative agreement with experiment, and elucidates key differences between a quantum dot and a QPC as a proximal charge sensor. Reduced screening and smaller characteristic energy needed to change transmission in the quantum dot compared to the QPC are responsible for its improved performance. 

\begin{acknowledgments}
We acknowledge funding from ARO/iARPA, the Department of Defense and
IBM. MK acknowledges support from Augustinus Foundation, H\o{}jgaard
Foundation and Stefan Rozental and Hanna Kobylinski Rozental
Foundation. This work was performed in part at the Center for
Nanoscale Systems (CNS), a member of the National Nanotechnology
Infrastructure Network (NNIN), which is supported by the National
Science Foundation under NSF award no. ECS-0335765. Computational
support from the NNIN computation project (NNIN/C) is gratefully
acknowledged. We thank Hendrik Bluhm, Edward Laird and David Reilly for useful discussion.
\end{acknowledgments}

\appendix


\begin{thebibliography}{14}
\bibitem[{field()}]{tsstepping}
The separation time  $\tau_{\rm{S}}$ is stepped from 1 to 200 ns, every 1000 cycles
for a total of 30,000 cycles.
\bibitem[{field()}]{Singleshotparams}
For QPC1 in Ref.~\cite{Singleshotpaper}, the fit parameters for the histograms, are $T_1 =34~\mu{\rm{s}}$,  $\tau_0=700$~ns, and $\langle P_{T}\rangle \sim 0.5$.
\bibitem{Johnsonnoise} The contribution from thermal fluctuations to intrinsic sensor noise is negligible. The cryogenic amplifier contributes $60\%$ ($40\%$) of the total noise for SQD (QPC1).
\end{thebibliography}

\end{document}